\documentclass[twocolumn,superscriptaddress,showpacs,amsmath,footinbib,prl]{revtex4-1}
\usepackage{amsmath}
\usepackage{mathtools} 
\usepackage{graphicx}
\usepackage{dcolumn}
\usepackage{bm}
\usepackage{hyperref}
\hypersetup{
	bookmarks=true,
	colorlinks=true,
	urlcolor=blue,
	citecolor=blue,
	linkcolor=blue, 
	pdftex,
	linktocpage=true, 
	linktoc=all,     
	hyperindex=true
}
\usepackage[mathlines]{lineno}
\usepackage{siunitx} 
\usepackage{comment} 
\usepackage{braket} 
\usepackage{xcolor}

\usepackage{newfloat}
\DeclareFloatingEnvironment[name={Supplementary Figure},fileext=lsf,listname={List of Supplementary Figures}]{suppfigure}

\usepackage[normalem]{ulem}
\newcommand\redout{\bgroup\markoverwith
	{\textcolor{red}{\rule[.5ex]{2pt}{0.4pt}}}\ULon}

\begin{document}
\title{Supersolidity in Two-Dimensional Trapped Dipolar Droplet Arrays}
\author{J. Hertkorn}%
\affiliation{5. Physikalisches Institut and Center for Integrated Quantum Science and Technology, Universit\"at Stuttgart, Pfaffenwaldring 57, 70569 Stuttgart, Germany
}%
\author{J.-N. Schmidt}%
\affiliation{5. Physikalisches Institut and Center for Integrated Quantum Science and Technology, Universit\"at Stuttgart, Pfaffenwaldring 57, 70569 Stuttgart, Germany
}%
\author{M. Guo}%
\affiliation{5. Physikalisches Institut and Center for Integrated Quantum Science and Technology, Universit\"at Stuttgart, Pfaffenwaldring 57, 70569 Stuttgart, Germany
}%
\author{F. B\"ottcher}%
\affiliation{5. Physikalisches Institut and Center for Integrated Quantum Science and Technology, Universit\"at Stuttgart, Pfaffenwaldring 57, 70569 Stuttgart, Germany
}%
\author{K.S.H. Ng}%
\affiliation{5. Physikalisches Institut and Center for Integrated Quantum Science and Technology, Universit\"at Stuttgart, Pfaffenwaldring 57, 70569 Stuttgart, Germany
}%
\author{S.D. Graham}%
\affiliation{5. Physikalisches Institut and Center for Integrated Quantum Science and Technology, Universit\"at Stuttgart, Pfaffenwaldring 57, 70569 Stuttgart, Germany
}%
\author{P. Uerlings}%
\affiliation{5. Physikalisches Institut and Center for Integrated Quantum Science and Technology, Universit\"at Stuttgart, Pfaffenwaldring 57, 70569 Stuttgart, Germany
}%
\author{H.P. B\"uchler}%
\affiliation{Institute for Theoretical Physics III and Center for Integrated Quantum Science and Technology, Universit\"at Stuttgart, Pfaffenwaldring 57, 70569 Stuttgart, Germany
}%
\author{T. Langen}%
\affiliation{5. Physikalisches Institut and Center for Integrated Quantum Science and Technology, Universit\"at Stuttgart, Pfaffenwaldring 57, 70569 Stuttgart, Germany
}%
\author{M. Zwierlein}%
\affiliation{MIT-Harvard Center for Ultracold Atoms, Research Laboratory of Electronics, and Department of Physics, Massachusetts Institute of Technology, Cambridge, Massachusetts 02139, USA}
\author{T. Pfau}%
\email{t.pfau@physik.uni-stuttgart.de}
\affiliation{5. Physikalisches Institut and Center for Integrated Quantum Science and Technology, Universit\"at Stuttgart, Pfaffenwaldring 57, 70569 Stuttgart, Germany
}%

\date{\today}
\begin{abstract}
	We theoretically investigate the ground states and the spectrum of elementary excitations across the superfluid to droplet crystallization transition of an oblate dipolar Bose-Einstein condensate. We systematically identify regimes where spontaneous rotational symmetry breaking leads to the emergence of a supersolid phase with characteristic collective excitations, such as the Higgs amplitude mode. Furthermore, we study the dynamics across the transition and show how these supersolids can be realized with standard protocols in state-of-the-art experiments.
\end{abstract}

\maketitle

Supersolids are states of matter that simultaneously combine the frictionless flow of superfluids with the crystalline order of solids \cite{Prokofev2007,Balibar2010,Boninsegni2012}. Their existence was initially proposed with macroscopic systems in mind, such as superfluid helium undergoing a transition to a solid  \cite{Gross1957,Andreev1969,Thouless1969,Chester1970,Leggett1970}.
Despite decade-long efforts, conclusive experimental evidence has been elusive in these systems \cite{Chan2013}. In contrast, supersolids have been experimentally realized in ultracold atomic gases with strong magnetic dipolar interactions \cite{Tanzi2019, Bottcher2019, Chomaz2019, Guo2019, Tanzi2019a, Natale2019, Bottcher2020, Hertkorn2021}.
In spite of rapid developments in this field \cite{Roccuzzo2019,Tanzi2021,Hertkorn2019,Bottcher2020, Blakie2020supersolelongate, Hertkorn2021,Blakie2020variational,Pal2020numberfluct,Chomaz2020whatCanBeLearnt,Ilzhoefer2021}, the dipolar supersolids have so far been experimentally limited to cigar-shaped geometries. In this setting, dipolar Bose-Einstein condensates (BECs) give rise to one-dimensional (1D) crystals of overlapping quantum droplets, stabilized by quantum fluctuations \cite{Bulgac2002,Schutzhold2006,Lima2011,Lima2012,Petrov2015,Ferrier-Barbut2016}. The density overlap facilitates superfluid flow throughout the crystal, giving rise to the supersolid droplet (SSD) phase. The SSD formation itself is driven by a roton instability \cite{Hertkorn2021}. This instability is a consequence of the rotonic excitation spectrum of dipolar BECs, which is reminiscent of the spectrum of superfluid helium \cite{Feynman1954heTwoFluid,Santos2003,Chomaz2018,Hertkorn2021}.

Extending the dipolar supersolids to two-dimensional (2D) crystal structures is an imperative step toward the long-standing goal of realizing macroscopic supersolid structures \cite{Boninsegni2012,Saccani2012,Cinti2014,Kora2019}. Theoretically, the phase diagram of 2D dipolar supersolids has been discussed in an infinite system \cite{Zhang2019} and up to now studies in finite-size systems were limited to high atom numbers and specific choices of trap parameters \cite{Baillie2018,Roccuzzo2020,Gallemi2020}. A priori, it is unclear which parameters are generally favorable for the formation of 2D SSDs and whether they can be experimentally realized. First experimental steps toward 2D SSDs have recently been made \cite{Schmidt2021,Norcia2021}.

Here, we show how 2D SSDs emerge from superfluid BEC states in oblate harmonic trap geometries. First, we note that general scaling properties of dipolar BECs can be used to effectively modify the strength of the stabilizing quantum fluctuations, which simplifies the identification of regimes in which a large overlap between crystallizing droplets is maintained. Second, we investigate the excitation spectrum across the superfluid to supersolid phase transition. We show that the underlying ground states (GSs) exhibit unique supersolid properties by finding elementary excitations characteristic of the supersolid phase, namely low-energy Goldstone and Higgs amplitude modes. These modes arise through spontaneous symmetry breaking of the continuous rotational symmetry of the BEC. Third, we show that these SSDs can be dynamically reached with an experimentally feasible protocol. 

At zero temperature, an effective mean-field description of dipolar BECs is provided by the extended Gross-Pitaevskii equation (eGPE) \cite{Ronen2006,Saito2016,Ferrier-Barbut2016,Wenzel2017,Roccuzzo2019}. In this description, quantum fluctuations are included as beyond-mean field effects according to the Lee-Huang-Yang correction for dipolar systems \cite{Schutzhold2006,Lima2011,Lima2012,Petrov2015,Ferrier-Barbut2016}. Contact and dipolar interaction strengths between $N$ atoms are controlled by the scattering length $a_s$ and dipolar length $a_\mathrm{dd} = \mu_0 \mu_m^2 M / 12 \pi \hbar^2$ with the magnetic moment $\mu_m$ and the mass $M$, respectively. For the full eGPE, we refer to the supplementary material \cite{supmat}. Within this framework, we find the GS $\psi_0$ of the system using gradient flow \cite{Bao2010} and conjugate gradient \cite{Modugno2003,Ronen2006,Antoine2017,Antoine2018} techniques. On top of this GS, we study elementary excitations with the Bogoliubov-de Gennes (BdG) formalism, extensively used to understand 1D SSDs \cite{Roccuzzo2019,Guo2019,Natale2019,Hertkorn2019,Hertkorn2021} and described in detail in Refs.~\cite{Ronen2006,Baillie2017,Chomaz2018,Hertkorn2019}. In short, we linearly expand the wavefunction ${\psi(\boldsymbol{r},t) = \left\{ \psi_0(\boldsymbol{r}) + \alpha \left[ u(\boldsymbol{r}) e ^{-i\omega t} + v^{\ast}(\boldsymbol{r})e^{i\omega t} \right] \right\} e^{-i\mu t/\hbar}}$ with a small amplitude $\alpha$ around the GS with chemical potential $\mu$. This ansatz leads to a system of linear equations, the BdG equations \cite{Baillie2017,Chomaz2018,Hertkorn2019}. We numerically solve these equations and obtain the Bogoliubov amplitudes $u_i,\,v_i$ corresponding to the lowest excitation energies $\hbar \omega_i$ and the functions $f_i = u_i + v_i$ related to density fluctuation patterns, which we use to characterize the modes. Due to our finite-size system, we obtain a discrete spectrum of elementary excitations. For further details on the methods, we refer to Refs.~\cite{supmat,Hertkorn2019}.

In the following, we will be most interested in parameter regimes where the GS of a dipolar BEC transitions to the elusive SSD phase as the scattering length $a_s$ is tuned. To this end, we notice scaling relations of dipolar BECs in the presence of quantum fluctuations. In a cylindrically symmetric trap ${V_\mathrm{ext}(\boldsymbol{r}) = M \omega_0^2 (x^2 + y^2 + \lambda^2 z^2)/2}$ with aspect ratio $\lambda$, dimensionless units based on the length scale ${x_s = \sqrt{\hbar/M \omega_0}}$ and a wavefunction normalized to unity \cite{Bao2013,supmat} reveal that the mean-field interactions are controlled by the dimensionless contact interaction strength $C \propto a_s N / x_s$ and dimensionless dipolar interaction strength $D \propto a_\mathrm{dd} N / x_s$ \cite{Ronen2006,Blakie2012}. Whereas the linear atom number density $N/x_s$ influences the strength of both mean-field interactions in exactly the same way, the dimensionless quantum fluctuation strength scales as $Q \propto C^{5/2}/N$ \cite{supmat}. While holding $N/x_s$ constant, higher trapping frequencies and, notably, lower atom numbers enhance the quantum fluctuations. The enhanced quantum fluctuations generally lead to a stronger stabilization and higher overlap of the droplet array. Therefore, the scaling enables to effectively tune the quantum fluctuations and allows for the study of supersolids at lower atom numbers than previously considered \cite{Baillie2018,Roccuzzo2020,Gallemi2020}. 

\begin{figure}[tb!]
	\includegraphics[trim=0 0 0 0,clip,scale=0.55]{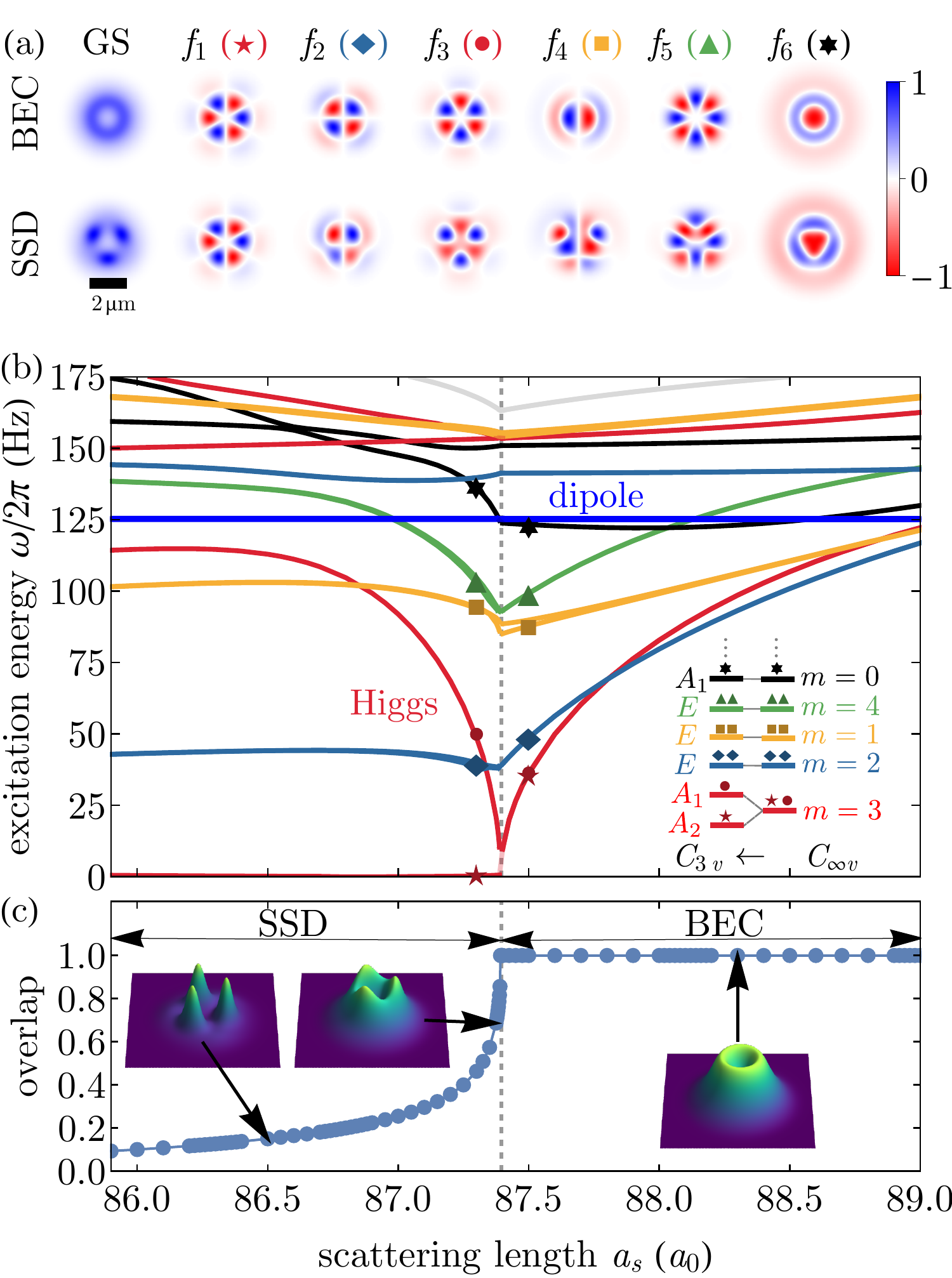}
	\caption{(a) GS density and low-lying density fluctuation mode patterns $f_i$ in the BEC (top row) and on the SSD side (bottom row) close to the transition. The modes on the BEC side can be characterized as radial ($f_6,\, m=0$) and angular $(f_1$-$f_5,\, m>0)$ roton modes, where $m$ is the number of angular nodal lines. The supersolid inherits modes from the BEC, but with a spontaneously broken rotational symmetry, giving rise to characteristic excitations such as the Higgs amplitude mode ($f_3$). The modes $(f_1$-$f_6)$ are numbered according to their energies on the droplet side (see (b)). The images are normalized to 1 for visual clarity (the colorbar applies to all images in arbitrary units). (b) Excitation energies of the low-lying modes as a function of the scattering length $a_s$ around the transition point. The $m=3$ angular roton modes trigger the phase transition at the critical scattering length $a_{s,c} \simeq 87.35\,a_0$ (dashed vertical line) to a supersolid droplet crystal, which spontaneously breaks the continuous rotational symmetry ($C_{\infty v}$) to a discrete threefold rotational symmetry ($C_{3v}$). In the supersolid, modes are labelled according to the irreducible representation of the remaining symmetry group (here for $C_{3v}$: $A_1,\,A_2,\,E$). (c) GSs and droplet density overlap as a function of $a_s$. In the BEC, the overlap is set to unity. The points show the sampling of the GSs.}
	\label{fig:fig1}
\end{figure}

We study a strongly dipolar BEC of ${N=20\times 10^3}$ magnetic $^{162}$Dy atoms (${a_\mathrm{dd} \simeq 130\, a_0}$) confined in an oblate harmonic trap with trapping frequencies ${\omega/2\pi = (125,\,125.5,\,250)\,\si{\hertz}}$, aspect ratio $\lambda \simeq 2$ and a magnetic field pointing along $\hat{\boldsymbol{z}}$. A small trap asymmetry of $0.4\,\si{\percent}$ is included to provide a slight preference in orientation for the collective modes \cite{Gallemi2020,NoteDetailsAsymm}. We map out the GSs and the collective excitation spectrum across the superfluid to supersolid phase transition as a function of $a_s$. Figure.~\ref{fig:fig1}(a) shows relevant BdG excitations in the vicinity of the quantum critical point. Figure~\ref{fig:fig1}(b)-(c) show the behavior of the corresponding excitation energies and the GS distributions across the BEC-SSD transition. The chosen parameters allow us to focus on the most elementary supersolid with a 2D crystal structure, namely three droplets that self-organize in a triangular array (Fig.~\ref{fig:fig1}(c)). The scaling relations enable the realization of this elementary supersolid at experimentally feasible atom numbers and trapping frequencies \cite{supmat}.

Close to the transition to a droplet crystal, the usual Thomas-Fermi or parabola-shaped BEC GS develops into a biconcave blood cell-like shape, as shown in Fig.~\ref{fig:fig1}(c). Qualitatively, one can understand the emergence of blood cell-shaped GSs since the dipoles in oblate traps predominantly exert repulsive long-range forces in the weakly confined directions and push part of the density to the outer rim  \cite{Eberlein2005,Ronen2006,Ronen2007}. Three droplets emerge on the ring of the blood cell for $a_s < a_{s,c} \simeq 87.35\,a_0$ (dashed line in Fig.~\ref{fig:fig1}) and form an equilateral triangle with a high density overlap between the droplets. The overlap is defined as the ratio between minimum and maximum density along the ring on which the droplets form and is a simple measure of the superfluid fraction \cite{Bottcher2019,Hertkorn2019,Zhang2019,Aftalion2007}. The relatively small jump of the droplet overlap at the transition (Fig.~\ref{fig:fig1}(c)) is consistent with a first-order quantum phase transition \cite{Saccani2012,Macri2013,Lu2015,Zhang2019,Roccuzzo2020}. Collective excitations on top of the blood cell GSs are enhanced toward the transition, which makes them fragile objects and has precluded a direct experimental observation so far. However, recently indirect evidence of the blood cell shapes has been found \cite{Schmidt2021}. Their fragility and the formation of three droplets in the GS are closely linked by the excitation spectrum of blood cell GSs for $a_s > a_{s,c}$ as we will show in the following.

The excitation spectrum of blood cell GSs features rotonic excitation modes, which can be classified as radial and angular roton modes (Fig.~\ref{fig:fig1}(a)-(b)) \cite{Ronen2006,Ronen2007,Wilson2008,Bisset2013,Schmidt2021}. While the sloshing (dipole) mode at the trap frequency of $125\,\si{Hz}$ is the lowest-lying mode on the BEC side at large scattering lengths and unaffected by a change of scattering length \cite{Kohn1961,Ronen2006,Hertkorn2019}, the roton modes soften below the dipole mode as the biconcave contrast increases toward $a_{s,c}$. Radial roton modes (Fig.~\ref{fig:fig1}(a), $f_6$) are cylindrically symmetric and have ring-like density modulations at a finite radial wavevector. Angular rotons (Fig.~\ref{fig:fig1}(a), $f_1$-$f_5$) have an additional angular oscillatory structure \cite{Ronen2007,Schmidt2021}. With the underlying blood cell GS, angular rotons can be understood as a circular density modulation on top of the density ring \cite{Ronen2007,Schmidt2021}. These modes are characterized by the integer $m > 0$ and are twofold degenerate due to the combination of the rotational symmetry around the $z$-axis and the mirror symmetry with respect to vertical planes through the trap center and parallel to the $z$-axis. The rotational symmetry allows a characterization of the modes by their angular momentum $\pm m$ along the $z$-axis \cite{Ronen2006,Wilson2008,Wilson2009,Wilson2009AngularCollapse,Bisset2013}, while the mirror symmetry guarantees the degeneracy of the two angular momentum states $\pm m$, i.e. the modes carry an irreducible representation of the symmetry group $C_{\infty v}$ \cite{Dresselhaus2008}. In particular, the two degenerate modes have the behavior $\sin(m \phi)$ and $\cos(m \phi)$, respectively, where $\phi$ is the azimuthal angle \cite{RotonOrientation,Ronen2006,Wilson2008,Wilson2009}. The weak splitting of the modes with $m=1$ close to the transition is a consequence of the slight asymmetry of the trap \cite{Schmidt2021}.

The biconcave shape of the GS facilitates an angular mean-field instability when one of the angular roton modes is sufficiently low or becomes imaginary. As one can see in Fig.~\ref{fig:fig1}(b), the various roton modes soften with different rates as the scattering length is reduced, changing the ordering of the lowest-lying modes as the point of instability is approached. Intuitively, the symmetry of the lowest-lying mode at the instability point gives rise to a crystal structure of the same symmetry. A nontrivial interplay between trap aspect ratio and interaction strengths determines which mode is the lowest at the transition \cite{Wilson2009}. In the present case, the two degenerate ${m=3}$ angular roton modes (Fig.~\ref{fig:fig1}(a), $f_1$ and $f_3$) soften fastest close to $a_{s,c}$ and give rise to three droplets (Fig.~\ref{fig:fig1}(c)). This is analogous to the softening of two degenerate linear roton modes in cigar-shaped traps \cite{Guo2019,Natale2019,Hertkorn2019,Hertkorn2021}, leading to the formation of 1D SSDs. However, a key difference between 1D and 2D SSD formation is the symmetry giving rise to the SSDs. While in elongated systems the translational symmetry is already broken due to the presence of the external trap \cite{Tanzi2019,Bottcher2019,Chomaz2019,Norcia2021,supmat}, 2D SSDs arise from the breaking of a genuinely continuous (rotational) symmetry, despite the presence of a trap. For further comparisons between 1D and 2D SSDs, we refer to the supplementary material \cite{supmat}.

When three droplets form, the rotational symmetry is broken to a discrete threefold symmetry. The remaining symmetry group is $C_{3v}$ and the Bogoliubov modes are characterized by two one-dimensional irreducible  representations $A_1$ and $A_2$  (symmetric and antisymmetric under reflections, respectively and both symmetric under rotations by $2\pi/3$), and a two-dimensional irreducible representation $E$ \cite{Dresselhaus2008}. The modes of the BEC split at the transition to the supersolid according to the compatible irreducible representations of $C_{3v}$ as ${m=3+3k \rightarrow (A_1,\,A_2)}$ for any integer $k \geq 0$, ${m=0 \rightarrow A_1}$ and ${m=1,\,2,\,4,\,5,\,... \rightarrow E}$ (Fig.~\ref{fig:fig1}). The critical angular roton modes with $m=3$ are compatible with $A_1$ and $A_2$, and therefore split up in the supersolid phase (Fig.~\ref{fig:fig1}(b)). The mode compatible with $A_1$ fixes its nodes at the droplet positions (Fig.~\ref{fig:fig1}(a), $f_1$) and exhibits a very low energy determined by the asymmetry of the external trap. In contrast, the mode compatible with $A_2$ matches its spatial maxima to the droplet positions (Fig.~\ref{fig:fig1}(a), $f_3$) and rises quickly with energy. The spatial pattern of the former (Fig.~\ref{fig:fig1}(a), $f_1$), shows that this mode corresponds to a rotation of the droplet array. In a perfectly symmetric trap, this ``mode" accounts for the GS degeneracy of the spontaneously broken rotational symmetry in the supersolid phase and has zero excitation energy. The spatial pattern of the latter (Fig.~\ref{fig:fig1}(a), $f_3$), shows that the mode is an amplitude modulation between the droplet density and the superfluid background, which is the Higgs amplitude mode associated to the spatial symmetry breaking \cite{Endres2012,Leonard2017higgsgoldstone,Hertkorn2019}.

The Higgs mode on the droplet side of the transition quickly increases in energy and strongly hybridizes with higher modes, as seen in Fig.~\ref{fig:fig1}(b). In the BEC regime, only modes in the same $m$-subspace can couple to each other \cite{Neumann-Wigner1929,landau1981quantum,Ronen2006,Wilson2009,Bisset2013}, leading to avoided crossings and hybridization between angular roton modes and BEC phonon modes of the same $m$ \footnote{For example, the $m=2$ roton mode and the quadrupole mode have an avoided crossing, and similarly higher BEC phonon modes have avoided crossings with higher $m$ angular roton modes \cite{Bisset2013,Schmidt2021}}.  As $m$ ceases to be a good quantum number when the rotational symmetry is broken from the BEC to the SSD phase, the Higgs mode can couple to modes compatible with $A_1$ in the supersolid. Generally, the Higgs mode can couple to other modes as well, since the absence of Lorentz invariance in our system precludes the existence of pure phase and amplitude modes \cite{Bruun2014,Pekker2014,Leonard2017,Hertkorn2019}. The coupling of the Higgs mode in the higher branches is enhanced compared to the elongated supersolids \cite{Hertkorn2019}. Observing the hybridization of the Higgs mode and strong increase in energy for these higher collective modes by techniques such as Bragg spectroscopy or a direct in-situ observation of the dynamics would provide an indication of supersolidity in oblate traps.

The next higher modes that can be seen in the spectrum (Fig.~\ref{fig:fig1}(a), $f_2$ and its degenerate partner) are translations along $x$- and $y$-directions, with a density adjustment that keeps the center of mass close to the trap center. This allows these modes to have a low energy in the droplet regime, analogous to the Goldstone mode in elongated traps \cite{Guo2019,supmat}.

While the above discussion is based on a specific trap geometry, we find the above observations to be true in various trap geometries and atom numbers and therefore consider them as general properties of the BEC to SSD phase transition. Namely, we find that the phase transition is driven by two degenerate roton modes, which split at the transition point into a low-energy mode and a Higgs mode that strongly hybridizes with higher-lying modes and as a result becomes strongly damped. For instance, in the same trap geometry and $N=50\times 10^3$, we find a transition from BEC to four droplets, facilitated by the softening of two degenerate $m=4$ roton modes. The remaining symmetry group after the spontaneous rotational symmetry breaking to four droplets is $C_{4v}$ \cite{Dresselhaus2008}, and the modes split to the compatible irreducible representations according to ${m=4+4k \rightarrow (A_1,\,A_2)}$, ${m=2+4k \rightarrow (B_1,\,B_2)}$ for any integer $k \geq 0$ and ${m=0 \rightarrow A_1}$, ${m=1,\,3,\,5,... \rightarrow E}$.

\begin{figure}[tb!]
	\includegraphics[trim=0 0 0 0,clip,scale=0.53]{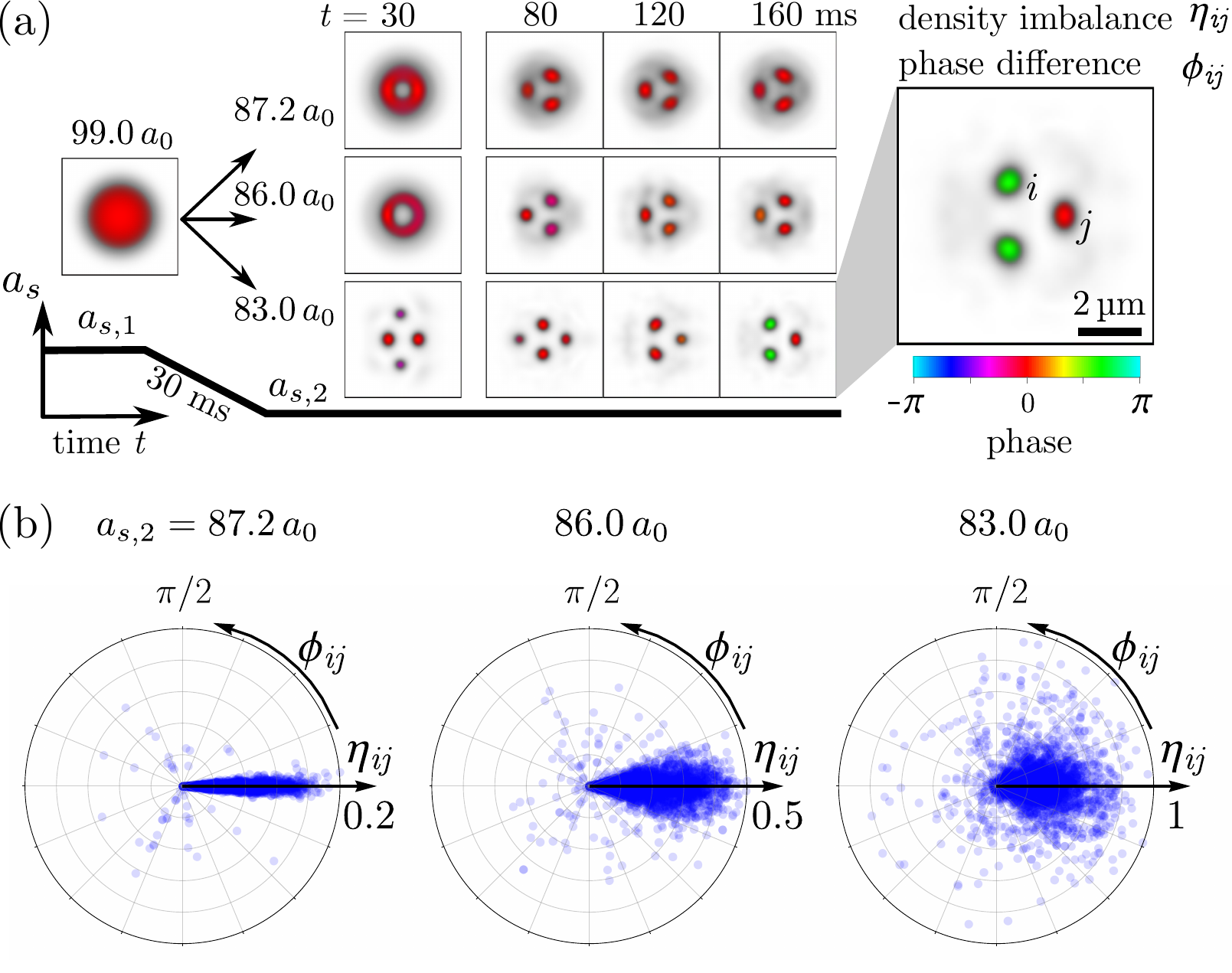}
	\caption{(a) Real-time evolution of the eGPE with a linear scattering length ramp, starting from a BEC at ${a_{s,1} = 99.0\,a_0}$ and ending on the other side of the transition at ${a_{s,2} = \{87.2,\,86.0,\,83.0\}\,a_0}$ after $30\,\si{\milli\second}$. The atom number is $N=20\times 10^3$. The dynamics are run 20 times with different initial noise, seeded at $t=0\,\si{\milli\second}$ (not shown). The images show one realization of the time evolution of 2D density cuts $n(x,y,0)$, color coded with the phase of the wavefunction. The droplet density imbalance $\eta_{ij}$ and phase difference $\phi_{ij}$ is evaluated for $t \geq 50\,\si{\milli\second}$ between all pairs $\{i,j\}$ of droplets (see (b)). (b) Imbalance $\eta_{ij}$ and phase difference $\phi_{ij}$ for $t \geq 50\,\si{\milli\second}$. Ramping the scattering length to final values in the vicinity of the transition ($\{87.2,\,86.0\}\,a_0$) allows the droplets to form with a balanced density and relatively high overlap, which maintains their global phase coherence. Ramping far away from the transition ($83.0\,a_0$) leads to strongly imbalanced droplets in many realizations, which lose their phase coherence due to a small overlap between the droplets in some realizations. The radii of the polar plots indicating the imbalance are scaled to maintain visibility of $\phi_{ij}$ even for the smaller imbalances at $a_{s,2} = \{87.2,\,86.0\}\,a_0$ compared to $a_{s,2} = 83.0\,a_0$.}
	\label{fig:fig2}
\end{figure}
To investigate whether the formation of 2D SSDs is experimentally feasible, we perform real-time simulations of the eGPE following a standard experimental procedure as shown in Fig.~\ref{fig:fig2}(a). We consider ${N=20\times 10^3}$ atoms (cf. Fig.~\ref{fig:fig1}). Starting from the GS in the BEC regime at ${a_{s,1} = 99.0\,a_0}$, we ramp the scattering length linearly in $30\,\si{\milli\second}$ to final scattering lengths ${a_{s,2} = \{87.2,\,86.0,\,83.0\}\,a_0}$ on the other side of the transition. In order to include possible experimental imperfections, we seed the initial dynamics with thermal noise corresponding to a temperature of $20\,\si{\nano\kelvin}$ according to the truncated Wigner description \cite{Blakie2008,Blakie2015,Blakie2016,Tanzi2019,Chomaz2019,Natale2019,Smith2021}. We repeat the simulation $20$ times with different random noise added to acquire representative dynamics of multiple experimental realizations. After roughly $t = 40\, \si{\milli\second}$ ($10\, \si{\milli\second}$ after the ramp is complete) the droplets have formed in all realizations. We evaluate the phase difference $\phi_{ij} = \phi_i - \phi_j$ and the density imbalance ${\eta_{ij} = |n_i - n_j|/\max(|n_i - n_j|)}$ for $t \geq 50\, \si{\milli\second}$ between all pairs of droplets $\{i,j\}$ (Fig.~\ref{fig:fig2}(b)). This allows us to monitor how strongly excited the droplets are after their formation, whether they lose their phase coherence, and if they approach the GS. Note that strongly imbalanced or incoherent droplets are far away from the GS \cite{Bottcher2019}. 

Figure.~\ref{fig:fig2}(a) shows that close to the transition (${a_{s,2} = 87.2\,a_0}$) where the droplets have a high overlap, the droplet formation process is smooth, all droplets form at the same time with a minimal imbalance, and the formation process does not lead to a loss of phase coherence (Fig.~\ref{fig:fig2}(b)). The GS density is approached and the uniform phase is maintained throughout the transition for every realization. When the scattering length is ramped further away from the transition point ($a_{s,2} = 86.0\,a_0$), the droplet formation process leads to stronger excitations of the droplet crystal and consequently a larger imbalance, seen in Fig.~\ref{fig:fig2}(b). However, the phase difference between droplets for all realizations and almost all times is small and roughly bounded by $|\phi_{ij}| \lesssim \pi/8$ \cite{supmat}. The droplets exhibit coherent dynamics after their formation, but do not lose their global phase coherence and therefore still represent a supersolid state. The phase rigidity of the supersolid around this scattering length and above is facilitated by a sufficiently high droplet overlap, maintaining a coherent exchange of atoms between the droplets.

As the scattering length is ramped further in the droplet regime, the droplet formation becomes more violent, which can be seen from the large density imbalance shown in Fig.~\ref{fig:fig2}(b) for $a_{s,2} = 83.0\,a_0$. In contrast to  the simulations for $a_{s,2} \gtrsim 86.0\,a_0$, the droplets dephase during the dynamics and the phase difference becomes more randomly distributed. In many realizations the droplets are not connected by a sufficiently high droplet overlap to support the stabilization of their relative phase in this case. The droplets form an isolated and incoherent crystal.

Figure~\ref{fig:fig2} illustrates that the supersolid GS can be dynamically approached in a narrow region of interaction strengths at atom numbers in reach of current state-of-the-art experiments \cite{Schmidt2021,Norcia2021}. Close to the transition, the forming droplets maintain their phase coherence, despite their residual excitations.

In conclusion, our work provides a comprehensive framework for the making, probing and understanding of two-dimensional dipolar supersolids. Together with the results of our dynamical simulations and the reported scaling relations, we outline a clear pathway for the realization of these states in experiments. We suggest that the coupling and hybridization of the Higgs amplitude mode with higher branches can serve as an experimental signature for supersolidity in future experiments seeking to realize supersolid two-dimensional dipolar droplet arrays \cite{Hertkorn2019}.

\section*{Acknowledgments}
M.G. and M.Z. acknowledge funding from the Alexander von Humboldt Foundation. T.L. acknowledges funding from the European Research Council (ERC) under the European Union’s Horizon 2020 research and innovation programme (Grant agreement No. 949431). This work is supported by the German Research Foundation (DFG) within FOR2247 under Pf381/16-1 and Bu2247/1, Pf381/20-1, FUGG INST41/1056-1 and the QUANT:ERA collaborative project MAQS.

\bibliography{refsC} 

\begin{thebibliography}{90}
\bibitem [{Not()}]{NoteDetailsAsymm}%
  \BibitemOpen
  \bibinfo {note} {{Using a slightly asymmetric trap is mainly convenient as
  almost degenerate modes will align their symmetry axes with the $x$- and
  $y$-axis, which allows them to be more easily distinguished and numerically
  analyzed. Also, we discuss a subtle difference between perfectly symmetric
  traps and how the spectrum is modified for a finite asymmetry later in the
  text.}}\BibitemShut {Stop}%
\bibitem [{Rot()}]{RotonOrientation}%
  \BibitemOpen
  \bibinfo {note} {{Angular rotons have the behavior $\propto \sin(m\phi +
  \phi_0)$, where $\phi_0$ is an arbitrary angle if the trap is perfectly
  cylindrically symmetric. For slight asymmetries in the trap, for example
  along the $x$- and $y$-axis, the degeneracy is lifted and the modes orient
  along the principal axes of the trap. Then, the relative angular phase
  difference between two angular roton modes that are degenerate in the
  perfectly symmetric trap is $\phi_0 = \pi/2$, leading to the behavior
  $\propto \sin(m\phi)$ and $\propto \cos(m\phi)$, respectively}.}\BibitemShut
  {Stop}%
\end{thebibliography}

\clearpage

\setcounter{figure}{0}
\renewcommand{\figurename}{Supplementary Figure}
\renewcommand{\thefigure}{S\arabic{figure}} 
\renewcommand{\theequation}{S.\arabic{equation}}
\newcounter{SFfig}
\renewcommand{\theSFfig}{S\arabic{SFfig}}

\section{Supplementary material}
\subsection{Numerical methods and BdG spectrum}\label{sec:nummeth}
Our effective mean-field description of dipolar Bose-Einstein condenstates (BECs) is based on the zero-temperature theory provided by the extended Gross-Pitaevskii equation (eGPE) \cite{Ronen2006,Wenzel2017,Roccuzzo2019} ${i \hbar \partial_t \psi = \hat{H} \psi}$, where $\psi$ is normalized to the atom number ${N=\int \mathrm{d}^3r\, |\psi(\boldsymbol{r},t)|^2}$ and ${\hat{H} \coloneqq \hat{H}_0 + g_s|\psi|^2 + g_\mathrm{dd} (U_\mathrm{dd} * |\psi|^2) + g_\mathrm{qf} |\psi|^3}$. The term ${\hat{H}_0 = -\hbar^2 \nabla^2 / 2M + V_\mathrm{ext}(\boldsymbol{r})}$ contains the kinetic energy and trap confinement ${V_\mathrm{ext}(\boldsymbol{r}) = M(\omega_x^2 x^2 + \omega_y^2 y^2 + \omega_z^2 z^2)/2}$, where $M$ is the mass.  The scattering length $a_s$ controls the contact interaction strength ${g_s = 4\pi\hbar^2a_s/M}$ and the  dipolar length $a_\mathrm{dd} = \mu_0 \mu_m^2 M / 12 \pi \hbar^2$ with the magnetic moment $\mu_m$ controls the dipolar interaction strength $g_\mathrm{dd} = 4\pi\hbar^2a_\mathrm{dd}/M$. The dipolar mean-field potential is given by the convolution ${(U_\mathrm{dd} * |\psi|^2)(\boldsymbol{r},t) = \int \mathrm{d}^3r'\, U_\mathrm{dd}(\boldsymbol{r}-\boldsymbol{r}') |\psi(\boldsymbol{r}', t)|^2}$, where $U_\mathrm{dd}(\boldsymbol{r}) = (3/4\pi) (1-3\cos^2 (\theta))/|\boldsymbol{r}|^3$ is the dipolar interaction for dipoles aligned along the magnetic field direction ($\hat{\boldsymbol{z}}$) and $\theta$ is the angle between $\hat{\boldsymbol{z}}$ and $\boldsymbol{r}$. Quantum fluctuations within the local density approximation for dipolar systems \cite{Schutzhold2006,Lima2011,Lima2012,Petrov2015,Ferrier-Barbut2016} are included with the term $g_\mathrm{qf}|\psi|^3$, where ${g_\mathrm{qf} = (32/3\sqrt{\pi}) g_s a_s^{3/2} Q_5 (\epsilon_\mathrm{dd})}$ and $\epsilon_\mathrm{dd} = g_\mathrm{dd} / g_s = a_\mathrm{dd} / a_s$ is the relative dipolar strength.  In our simulations, we use the approximation  ${Q_5(\epsilon_\mathrm{dd}) \simeq 1+ 3\epsilon_\mathrm{dd}^2}/2$ \cite{Lima2012,Ferrier-Barbut2016,Wenzel2017,Bisset2016,Baillie2017}. We find ground states using gradient flow \cite{Bao2010} and conjugate gradient \cite{Modugno2003,Ronen2006,Antoine2017,Antoine2018} techniques adapted to include beyond mean-field effects. To evolve the eGPE in imaginary or real time, we use time splitting spectral techniques \cite{Antoine2013}. The mean-field dipolar potential is evaluated effectively using Fourier transforms, where we use a spherical cutoff for the dipolar interaction. The cutoff radius is set to the size of the simulation space such that there is no spurious interaction between periodic images \cite{Goral2002,Ronen2006,Lu2010}.

We study the elementary excitations within the Bogoliubov-de Gennes (BdG) formalism as described in Ref.~\cite{Hertkorn2019} and linearly expand the wavefunction ${\psi(\boldsymbol{r},t) = \left\{ \psi_0(\boldsymbol{r}) + \alpha \left[ u(\boldsymbol{r}) e ^{-i\omega t} + v^{\ast}(\boldsymbol{r})e^{i\omega t} \right] \right\} e^{-i\mu t/\hbar}}$ with a small amplitude $\alpha$ around the ground state $\psi_0$ with the chemical potential $\mu$. This ansatz together with the eGPE leads to a system of linear equations \cite{Baillie2017,Chomaz2018,Hertkorn2019}. We numerically solve these equations to obtain the modes $u_i$ and $v_i$ corresponding to the lowest excitation energies $\hbar \omega_i$ \cite{Ronen2006,Hertkorn2019}.  Due to our finite-size system, we obtain a discrete spectrum of elementary excitations. The density fluctuation patterns $2f_i \psi_0$ are related to the function $f_i = u_i + v_i$, which we use to characterize the modes in the main text \cite{Blakie2012,Martin2012,Jona-Lasinio2013,Hertkorn2021}.

The BdG modes in cylindrically symmetric traps for BEC and blood cell ground states can be classified as radial and angular roton modes \cite{Ronen2006,Ronen2007,Martin2012}. The angular roton modes with $m>0$ are twice degenerate \cite{Ronen2006,Wilson2009,Martin2012}. This degeneracy can be lifted by an asymmetry in the trap, which also causes the lowest-energy mode to stagnate at a finite energy instead of becoming gapless at the transition point, as well as lifting the Higgs gap to finite energy \cite{Hertkorn2019}. In practice, as we have chosen a slight asymmetry of $0.4\,\%$ in the radial trapping frequencies, the lowest-energy mode has a finite energy below or roughly equal to $0.5\,\si{Hz}$. We have confirmed that the energy of this mode close to the transition point decreases toward more symmetric traps.

We conclude this section by a comparison between the supersolid droplet (SSD) transition discussed in the main text and the crystallization transition observed in Ref.~\cite{Schmidt2021}. At the transition, the critical $m=3$ angular roton modes shown in Fig.~\ref{fig:fig1} in the main text are energetically separated by a few tens of $\si{\hertz}$ from the next higher modes, which provides a distinct length scale and symmetry on the SSD side. In contrast to the spectrum of this superfluid to SSD transition, the spectrum toward a pure crystallization transition observed in Ref.~\cite{Schmidt2021} showed two angular roton branches that become almost degenerate at the transition point. In this case, the two different symmetries of the angular roton modes lead to competing symmetries and length scales and complicate the GS structure on the SSD side. Furthermore the overlap for the 2D SSDs we consider here is high, as we tuned it using the scaling properties initially discussed in the main text and further below.

\subsection{Dynamic simulations}
We give some details on the evaluation of the density imbalance and phase difference here. To evaluate the density imbalance $\eta_{ij}$ and phase difference $\phi_{ij}$ between droplets after their formation, we use peak detection to find the location of the droplets at every time step and track the density and phase. We then calculate the pairwise density differences ${\delta n_{ij}  = n_i - n_j}$ and phase differences ${\phi_{ij} = \phi_i - \phi_j}$ between  all pairs of droplets ${\{i,\,j\}}$. We define the density imbalance shown in Fig.~\ref{fig:fig2} in the main text as ${\eta_{ij} = |\delta n_{ij}| / \max(|\delta n_{ij}|)}$, i.e. we normalize by the maximum density difference of all pairs and scattering lengths to provide the same axis for the radius in the polar plots of Fig.~\ref{fig:fig2}(b). In practice for the phase difference we use the density weighted average of the phases ${\phi_i \simeq \bar \phi_{i} = \int_{\Omega_i} \! n(x,y,0)\phi(x,y,0)\mathrm{d}^2r / \int_{\Omega_i} \!  n(x,y,0) \mathrm{d}^2r}$ in areas $\Omega_i$ around the density peak positions to avoid contributions of arbitrary phases where the density is close to zero, for example for isolated droplets.

To characterize the phase differences for the different scattering lengths, we mention two measures that can be used to qualitatively indicate the localization or randomness of the phases. First, one can consider the fraction $\chi$ of points being bounded by a certain threshold  $x \geq |\phi_{ij}|\ \mathrm{mod}\ 2\pi$. For the data shown in Fig.~\ref{fig:fig2}(b), one obtains for example ${\chi(x=\pi/8) \simeq \{99,\, 96,\, 75 \}\, \%}$ for ${a_{s,2} = \{87.2,\,86.0,\,83.0\}\,a_0}$. To cover $90\,\%$ of the points, i.e. to reach ${\chi = 0.9}$, one needs ${x \simeq \{ 0.20,\, 0.71,\, 2.04 \}\,\pi/8}$. Second, one can calculate the circular variance ${V = 1-R}$ \cite{Jammalamadaka2001,Ilzhoefer2021}, where ${R = \left| \sum_{k = 1}^{M} z_k \right|/M}$ with $z_k = e^{i\phi_k}$ ($\phi_{k}$ is $\phi_{ij}$ indexed linearly by $k=1,2,\ldots,M$). One obtains ${V \simeq \{1.2,\, 3.1,\, 12 \}\,\% }$ for ${a_{s,2} = \{87.2,\,86.0,\,83.0\}\,a_0}$. These three values indicate that global phase coherence is mostly maintained after ramping close to the transition. Ramping further into the isolated droplet regime reduces the degree of global phase coherence (Fig.~\ref{fig:fig2} in the main text).

We also evaluated the density difference to the ground state $d(t) = ||n(t) - n_\mathrm{GS}||^2/||n_\mathrm{GS}||^2$ as in Ref.~\cite{Bottcher2019}. We accounted for the arbitrary position of the droplet formation by rotating the images at every time step to match the ground state pattern (finding the angle for which the density difference is minimal), such that a lower bound on density difference to the ground state can be extracted. This quantity is mostly sensitive to the number of droplets. In almost all simulations, the density reaches a three-droplet state after $t \gtrsim 120\,\si{\milli\second}$, which is the same droplet number that the ground state has. Therefore the density difference reaches relatively small values and fluctuates between $0.2$ to $0.4$ in all cases, indicating that the density is relatively close to the ground state. Considering the droplet imbalance and phase difference then gives a more detailed picture whether droplets have lost their phase coherence and whether they show large imbalances during the dynamics.

\subsection{Reduced units and scaling properties}\label{sec:RedUnits}
Here we provide some additional information on the reduced units we used in the main text, in particular their relation to the interaction parameters $(a_s,\,a_\mathrm{dd},\,N)$ customarily used and how they may provide an intuition to connect similarities between different parameter regimes.

We define the dimensionless variables ${\tilde{t} = t\omega_0}$, ${\tilde{\boldsymbol{r}} =  \boldsymbol{r}/x_s}$, ${\tilde{\psi} = \psi \sqrt{x_s^{3}/N}}$ to nondimensionalize the eGPE  \cite{Lu2010,Blakie2012,Bao2013,Zhang2019,Lee2020}. Here, $\omega_0^{-1}$ and $x_s$ can at first be taken as arbitrary quantities with units of time and length, respectively. For Schr\"odinger-like equations it is convenient to define the energy and time units as ${\epsilon = \hbar^2 / M x_s^2}$ and ${\omega_0^{-1} = M x_s^2 / \hbar}$  based on the unit of length $x_s$, as this choice implies ${\epsilon \omega_0^{-1} = \hbar}$ and amounts to formally setting $\hbar = M = 1$ in the linear terms of the eGPE to obtain the dimensionless eGPE with rescaled coefficients for the nonlinear terms. Finally the dimensionless eGPE (omitting the tildes here and in the following) reads ${i \partial_t \psi = \hat{H} \psi}$, with ${\hat{H} = \hat{H}_0 + \hat{H}^{(\psi)}_\mathrm{int}}$. Here, ${\hat{H}_0 = -\nabla^2 / 2 + V_\mathrm{ext}(\boldsymbol{r})}$ and ${\hat{H}^{(\psi)}_\mathrm{int}(C,D,Q) \coloneqq  C |\psi|^2+ D (U_\mathrm{dd} * |\psi|^2) + Q |\psi|^{3}}$ are the linear and nonlinear parts of the eGPE, respectively. The dimensionless interaction strengths are given by
\begin{align}
C &= \frac{N g_s}{\hbar \omega_0 x_s^3} = \frac{4\pi a_s N}{x_s},\label{eq:supmatC} \\
D &= \frac{N g_\mathrm{dd}}{\hbar \omega_0 x_s^3} =  \frac{ 4\pi a_\mathrm{dd} N}{x_s} = \frac{\mu_0 \mu_m^2 M N}{3\hbar^2 x_s},\label{eq:supmatD} \\
Q &= \frac{N^{3/2}g_\mathrm{qf}}{\hbar \omega_0 x_s^{9/2}} = \frac{4}{3\pi^2} \frac{C^{5/2}}{N} Q_5(\epsilon_\mathrm{dd}),\label{eq:supmatQ}
\end{align}
where $\epsilon_\mathrm{dd} = D/C = g_\mathrm{dd} / g_s = a_\mathrm{dd} / a_s$.
In particular for harmonic external potentials it is useful to specifically let one of the trapping frequencies define the time unit by setting for example ${\omega_0 = \min_\alpha\{\omega_\alpha\}}$, yielding  ${V_\mathrm{ext}(\boldsymbol{r}) = \sum_\alpha \gamma_\alpha^2 x_\alpha^2 / 2}$ for ${\alpha \in \{x,\,y,\,z\}}$, where ${\gamma_\alpha = \omega_\alpha/ \omega_0}$. For cylindrically symmetric traps with ${\omega_x \simeq \omega_y \leq \omega_z}$, the external confinement is reduced to ${V_\mathrm{ext}(\boldsymbol{r}) \simeq (x^2 + y^2 + \lambda^2 z^2)/2}$, where ${\lambda = \gamma_z = \omega_z / \omega_0}$ is the aspect ratio. As we detailed in the main text, a significant consequence is that the contact and dipolar interaction terms are only ever modified by the product $N/x_s \propto N\sqrt{\omega_0}$ \cite{Goral2000}. For example, under the transformation $N \to s N$ and $\omega_0 \to \omega_0/s^2$, ${\hat{H}^{(\psi)}_\mathrm{int}(C,D,0)}$ is invariant for any $s$ while for the full interaction ${\hat{H}^{(\psi)}_\mathrm{int}(C,D,Q)}$, the $Q$-term behaves as $Q \to Q/s$.

\begin{figure}[tb!]
	\includegraphics[trim=0 0 0 0,clip,scale=0.49]{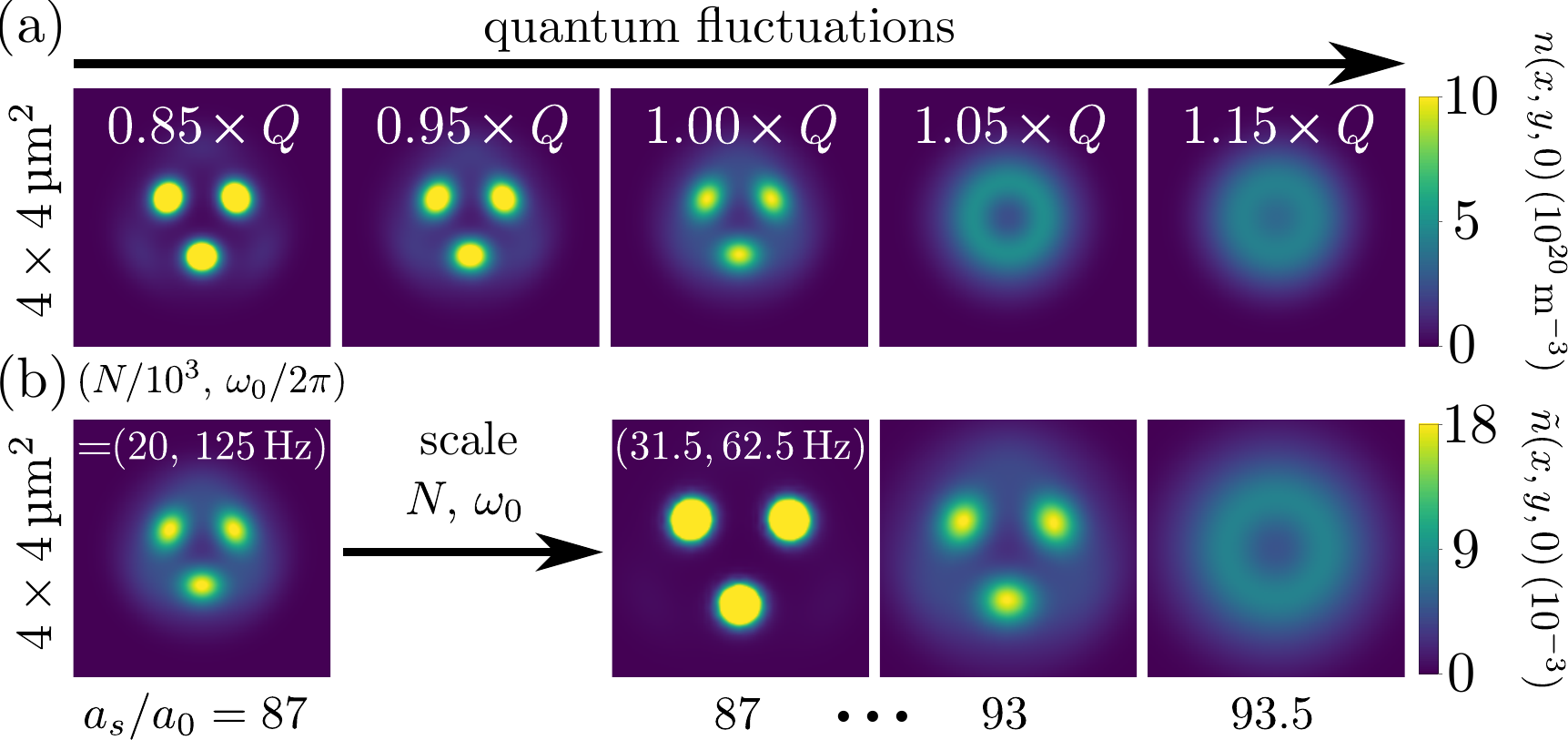}
	\caption{(a) Behavior of the ground state by varying the quantum fluctuations $Q$ by an arbitrary scale factor $s$, while leaving the contact and dipolar interaction strengths constant. This can be realized by using the scaling ${N\sqrt{\omega_0} = \mathrm{const.}}$~for different atom numbers $sN$ and trap geometries $\omega_0/s^2$. The ground state for $1.00\times Q$ was chosen with ${(N,\,\omega_0/2\pi,\,a_s) = (20 \times 10^3,\,125\,\si{\hertz},\,87\,a_0)}$ and recalculated for every modified $Q$. (b) Shows how the broken scale invariance can be used to relate similarities between different parameter regimes as an example for a trap frequency modified by a factor of two. The left image shows the ground state at $a_s = 87\,a_0$ (same as with $1.00\times Q$ in (a)). The density $n = |\psi|^2$ and the dimensionless density $\tilde{n} = n x_s^3/N$ at $z=0$ are shown shown in (a) and (b), respectively, to compare the ground states in both trapping geometries. The dimensionless peak densities of the states close to the transition (at ${\{87,\,93\}\,a_0}$) for both trap geometries are $\tilde{n}_0 \simeq 0.0183(2)$, corresponding to the peak densities $n_0 \simeq \{10.6,\,5.84\} \times \si{10\tothe{20}\,\meter\tothe{-3}}$. The ground state for ${(N,\,\omega_0/2\pi) = (31.5 \times 10^3,\, 62.5\,\si{\hertz})}$ reaches a similar overlap and dimensionless peak density $\tilde{n}_0$ near the transition ($a_s = 93\,a_0$) as the ground state in the higher confinement at lower atom numbers near its transition at $a_s = 87\,a_0$. Due to the effective reduction of $Q$, a higher scattering length and atom number are required to obtain a similar ground state, whereas two droplets are the ground state with only $\sqrt{2}N$ atoms at $a_s = 93\,a_0$.}
	\refstepcounter{SFfig}\label{fig:fig3}
\end{figure}

We illustrate these scaling arguments and their utility in Fig.~\ref{fig:fig3}. We take a three-droplet ground state (Fig.~\ref{fig:fig1}(a)) and vary the parameter $Q$ by a few percent to understand the effect of this scaling on the ground states. For smaller $Q$, the repulsive stabilization provided by the quantum fluctuations is decreased, and the relative strength of the dipolar interaction grows and more atoms accumulate in the attractive head-to-tail configuation in the droplets. For larger $Q$, the droplet overlap grows by an expulsion of atoms out of the droplets, facilitated by a stronger repulsive interaction until eventually, the blood cell and BEC states are recovered. Smaller (larger) $Q$ are achieved for lower (higher) trapping frequencies with correspondingly higher (lower) atom numbers in this picture. For the experimental search and study of supersolids in cylindrical traps, this means that higher confinements are desirable as they provide a way to effectively enhance the stabilizing quantum fluctuations at lower atom numbers. By considering higher trapping confinement (smaller $x_s$) one may, in view of a finite optical resolution, trade off the benefit of well-separated droplets (larger $x_s$) for equivalent supersolids at smaller atom numbers whose quantum fluctuations are enhanced.

Figure~\ref{fig:fig3}(b) shows this scaling by reducing the trapping frequencies by a factor of two while keeping the aspect ratio $\lambda = 2$ constant. We consider a ground state at $a_s = 87\,a_0$ on the left and expect to find a similar ground state for a factor of $\sqrt{2}$ higher atom number if there were no quantum fluctuations. Due to the change in quantum fluctuations, we actually need a slightly higher atom number than a factor of $\sqrt{2}$ (additional $11\,\si{\percent}$) to reach a three-droplet ground state in the new trap geometry. Despite the additionally provided atoms, the system in the lower confinement at the same scattering length $a_s = 87\,a_0$ is further in the isolated regime compared to the higher confinement. This is a consequence of the reduced quantum fluctuations in lower confinements, as can be seen by comparing Fig.~\ref{fig:fig3}(a) with the right hand side of Fig.~\ref{fig:fig3}(b). As a result, this scaling allows to efficiently find equivalent parameter regimes and to quickly locate favorable conditions for a BEC-supersolid phase transition. While here we only considered trap aspect ratios of two, these arguments are also valid for different cylindrically symmetric traps \cite{Bisset2016}.

We conclude with final remarks regarding the anology to the supersolids studied in elongated traps and behavior expected in asymmetric traps.

\subsection{Comparison of SSDs in cigar-, oblate-shaped and intermediate trap geometries}
We begin by comparing and contrasting supersolids in elongated and oblate traps. The formation of elongated SSDs is facilitated by two degenerate softening linear roton modes \cite{Guo2019,Hertkorn2019,Hertkorn2021}. These split at the transition into a low-energy Goldstone mode and a Higgs amplitude mode, analogous to the splitting of the two degenerate angular roton modes we find here in oblate traps. The degeneracy of the linear roton modes  in elongated traps arises due to an almost-flat density in the center of the trap, providing a quasi-continuous translational symmetry that is broken when the droplets form \cite{Hertkorn2019}.  However, strictly speaking the translational symmetry giving rise to elongated SSDs is already broken due to the presence of the external trap. In contrast, SSDs with 2D crystal structures arise from the breaking of a genuinely continuous (rotational) symmetry, despite the presence of a trap. Due to the continuity of the rotational symmetry that is broken in oblate traps, we obtain the mode associated with a rotation (Fig.~\ref{fig:fig1}(a), $f_1$) at arbitrarily low excitation energies determined by the asymmetry of the trap. The Goldstone mode in elongated traps is an out-of phase oscillation between the superfluid and crystal phonon and remains at low, but finite energies \cite{Guo2019,Hertkorn2019}. In both elongated and oblate geometries, the Higgs mode is an amplitude modulation between the superfluid background and the density in the droplets (Ref.~\cite{Hertkorn2019} and Fig.~\ref{fig:fig1}(a), $f_3$).

The intermediate regime between elongated and oblate trap geometries, where droplets are expected to arrange in a zigzag geometry similar to the structures known in ion crystals \cite{Fishman2008,Pyka2013}, can be intuitively understood from the two limiting cases (cylindrically symmetric trap in Fig.~\ref{fig:fig1} and elongated traps in Ref.~\cite{Hertkorn2019}) of the BdG spectrum. We know that in elongated geometries \cite{Hertkorn2019}, the supersolid arises because of the softening of two degenerate linear roton modes (softening on top of a BEC with a quasi-homogeneous density in the center of the trap, and mirror symmetry about its center along the long direction). Consider the behavior of a pair of degenerate angular roton modes in cylindrically symmetric trap geometries (Fig.~\ref{fig:fig1}), when the trap is slowly deformed to a more elongated geometry. As the asymmetry is increased, one of them rises and the other decreases -- they change shape and, roughly speaking, become less angular and more linear. One of them will eventually (continuously more so as the asymmetry increases) be associated to density fluctuations along the tight direction (the rising mode) and the other along the loose direction (the softening mode). Thus, when the asymmetry is intermediate, there are low-lying modes with structure both along the long and the short direction and the mode predominantly associated to density fluctuations along the long direction will eventually soften close to zero, triggering the phase transition to a zigzag pattern \cite{Fishman2008,Pyka2013}. Recently, this transition has been observed experimentally \cite{Norcia2021}.

\end{document}